\documentclass[aps,pre,reprint]{revtex4-2}
\usepackage{graphicx}

\begin{document}
\title{Percolation in a triangle on a square lattice}

\author{Zbigniew Koza}\email{zbigniew.koza@uwr.edu.pl}
\affiliation{Faculty of Physics and Astronomy, University of Wroc{\l}aw,
	50-204 Wroc{\l}aw, Poland}

\date{\today}

\begin{abstract}
Percolation on a plane is usually associated with clusters
spanning two opposite sides of a rectangular system.
Here we investigate three-leg clusters generated on a square lattice
and spanning the three sides of equilateral triangles.
If the position and orientation of the triangles relative to the lattice
are uniformly randomized, one obtains an efficient method
of determining the percolation threshold, on
par with the most advanced Monte Carlo methods developed
for the rectangular geometry. The universal crossing probability
for three-leg clusters is geometry-independent,
which opens a way for further improvements of the method.

\end{abstract}


\maketitle

\section{Introduction\label{sec::Introduction}}

With a simple, purely geometrical definition and complex behavior
that includes a phase transition, percolation
has become an important theoretical model in statistical physics.
It has also applications in various areas of science, like
conductivity in strongly heterogeneous solids \cite{Hunt2001,Besseaguet2019},
fluid flow in porous media \cite{Sahimi1993,Bolandtaba2011},
epidemics \cite{Ziff2021},
and thermal conductivity of composites~\cite{Shtein2015}.

While some critical exponents and crossing probabilities
in 2-dimensional systems
\cite{Stauffer1994,Cardy1992,Lawler2001,Smirnov2001ci,Flores2017},
as well as percolation thresholds in several particular models
\cite{Sykes1964,Kesten1980,Wierman2009} are known rigorously,
many results in this field has been obtained
with computer simulations. Over the years,  advanced
numercial methods have been developed, like the Leath
method~\cite{Leath1976,Lorenz1998,Xun2021},
hull-generating walks~\cite{Voss1984,Ziff1984},
gradient percolation~\cite{Rosso1985,Ziff1986,Tencer2021},
toroidal wrapping~\cite{Newman2001,Wang2013,Koza2016},
spanning clusters~\cite{Ziff92_PRL,Oliveira2003},
rescaled particles~\cite{Torquato12b},
frontier tracking~\cite{Quintanilla2000},
parallelized percolation on distributed machines \cite{Pruessner03},
dynamic programming \cite{Yang2013},
and the transfer matrix method \cite{Feng2008,Jacobsen2015}.
These methods, in all their diversity, share one common feature:
they are usually implemented with the assumption
that the system has a square (rectangular, hypercubic) geometry.
This choice is quite natural: it corresponds to
the most elementary and efficient computer data structure: array.
It also facilitates implementation of useful boundary conditions,
including wrapping boundaries. Wrapping boundaries, in turn,
enable one to study percolation on cylinders and tori,
the shapes for which strong theoretical results have been derived
and for which boundary effects are minimized, which results
in a faster convergence rate to the thermodynamic limit~\cite{Newman2001}.

The symmetry of a rectangle is closely related to the required connectedness
of a cluster: in spanning percolation the two sides that are checked for
spanning are geometrically equivalent, and so are the remaining two. One expects
that this configuration will produce quicker convergence to
the thermodynamic limit than, for instance, a trapezoid.
If one uses square systems, it is possible to investigate clusters
that span or wrap in both directions simultaneously \cite{Newman2001}.
But what about spanning in other number of directions than two or four,
for example, three?
It was recently argued that the probability, $p_3$ that,
in the thermodynamic limit,  there exists a three-leg cluster
touching the three sides of a triangle at the percolation threshold
has a universal value $1/2$ \cite{Koza2019,Flores2017}.
This result is supposed to be valid for any lattice and even systems
of arbitrary shape, in which case their perimeter must be divided into
three disjoint parts (arcs). For self-matching lattices this property
holds even for finite systems.

The aim of the paper is to investigate whether the property of
$p_3 \to 1/2$ can be used to develop an efficient
method of finding the percolation threshold, $p_\mathrm{c}$,
for planar lattices.
To this end, we investigate the well-known case of the site percolation
on a square lattice,  assuming, however, that the system
is in the shape of an equilateral triangle,
the simplest geometry with the three-fold symmetry corresponding
to the three legs of the clusters. The precise value of $p_\mathrm{c}$
in this model~\cite{Jacobsen2015},
\begin{equation}\label{eq::pc_Jacobsen}
p_\mathrm{c} = 0.592\,746\,050\,792\,10(2),
\end{equation}
will be used as a reference value against which the method
will be evaluated.


\section{Method\label{sec::Method}}
\begin{figure}
	\includegraphics[width=0.75\columnwidth]{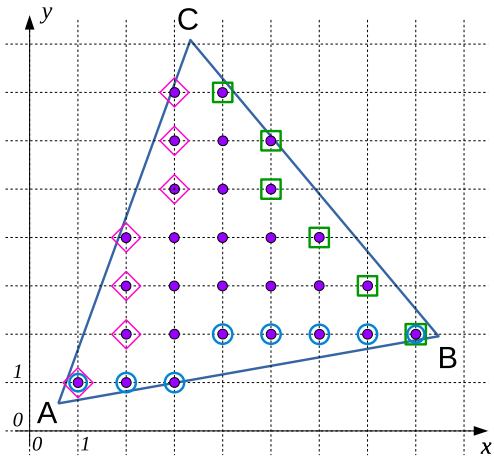}
	\caption{\label{fig:geometry} Geometry of the model. An
		equilateral triangle $ABC$ is located on a square lattice in such a way that
		its vertex $A$ is placed randomly inside $[0,1)\times[0,1)$ and the angle between
		side $AB$ and the direction of the $x$ axis is a random variable between 0 and 15 degrees.
		Circles, squares, and diamonds mark the "edge" site corresponding to three sides of the triangle.
		A three-leg percolating cluster must contain at least one site from each of these groups.
	}
\end{figure}

We start from  placing an equilateral triangle with sides of length $L$ on a square lattice
(Fig.~\ref{fig:geometry}).
Incompatibility  of the symmetries of the triangle and the lattice
gives rise to specific problems;
for example, the number of lattice sites encompassed by a triangle
is not invariant under translation and rotation, and the number of sites contained
inside a triangle of side length $L$ only approximately scales as $L^2$,
a problem particularly serious for small triangles.
There are also problems of topological nature: if one side of the triangle
is parallel to the $x$ axis, it cuts $\approx L$ lattice bonds, whereas
the remaining two sides cut $\approx \sqrt{3} L$ bonds. Thus, the sides,
even though of equal lengths, are not equivalent,
which may have an adverse effect on the simulation convergence rate.

To mitigate these problems, for each $L$ we consider an ensemble of
equilateral triangles randomly distributed and oriented relative to
the underlying lattice.
Symmetries of the lattice and the triangle allow to restrict the ensemble to
the cases where vertex $A$ is distributed uniformly inside the square
$[0,1)\times[]0,1)$, and the angle $\alpha$ made by side $AB$ and the direction
of the $x$ axis is distributed uniformly between 0 and 15 degrees.

Each simulation starts from randomization of the location
and orientation of the triangle. Then,
a minimum bounding box for the triangle is computed such that its
corners are  at lattice sites. This will be the arena of the simulation:
rectangular geometry trivializes the computation of the neighboring sites.
Next, the set of the sites encompassed by the triangle is determined.
We shall call them ``active sites''. Only active sites can be occupied
as the cluster grows.  Some of these sites are marked as
edge sites. We define three groups of edge sites, each corresponding
to a triangle side. A site is an edge site to side $AB$ if
a nearest-neighbor bond starting at this site cuts $AB$. Two other
groups are defined similarly for sides $BC$ and $CA$.
A site can be an edge site for more than one side
(for example, the lattice sites nearest to vertices
$A$ and $B$ in Fig.~\ref{fig:geometry}).
We define a three-leg percolating cluster as a cluster that contains
at least one site from each of these groups.

When the geometry has been established, a list of active sites is shuffled
randomly (we used the 64-bit Mersenne twister mt19937
random number generator from the standard C++ library).
During simulation proper, subsequent elements are popped from this list
and the corresponding sites are marked as occupied. With each new site
occupied, the union-find algorithm \cite{Newman2001,Cormen2009}
is used to monitor clusters' growth.
It also updates the information about the edge groups each cluster has reached.
To this end, before the simulation begins,
the union-find assigns to each site three bits that are used to signal
the property of being connected to a given group of edge sites.
This property is updated whenever clusters merge.
A simulation ends when, for some cluster,
all these bits are set to 1, which indicates that a three-leg
percolating cluster has just been formed.
Finally, the number of occupied sites is recorded.

In this way we obtained the distribution of the probability $R(n; L)$ that
for the ensemble of a randomly positioned and oriented
equilateral triangles of side $L$, with $n$ of
their internal sites occupied,
there is a three-leg cluster that spans all of its sides.
We introduce the occupation probability
\begin{equation} \label{eq::def}
	p = \frac{n}{S(L)},
\end{equation}
where $S(L) = (\sqrt{3}/4)L^2$ is the expected number of lattice sites contained inside
the triangle, which, due to the randomization of its placement and orientation,
is equal to its area. Clearly, at percolation, $0 < p < 1$.
Note, however, that in principle a triangle can hold more than $S(L)$ lattice
sites within itself, leading to the possibility of $p > 1$ when
every or nearly every site is occupied. This, however, cannot happen
at the onset of percolation; moreover,  the upper bound for $p$ as $L\to\infty$ is 1,
as it should.

An important quantity that we want to obtain from simulations is $R_L(p)$,
the probability that, for site occupancy $p$, a triangle of side $L$
contains a cluster spanning all of its sides.
While it is closely related to directly measured $R(n; L)$,
the relation is,
to some extent, unknown. First, we require $R_L(p)$
to be continuous, whereas $R(n; L)$ is discrete; second,
$R(n; L)$ is not known exactly, but is estimated from simulations.
The discrete nature of
$R(n; L)$ is especially problematic for small $L$. For example,
for $L=4$ only for 8 values of $n$ is this function different from 0 and 1.
The usual way of dealing with these problems is to fit the data
to some function (e.g. linear) in a vicinity of $p$.
Here, however, we use the canonical ensemble method \cite{Newman2001}.
Its main idea is to weight the values of a discrete-value function  $F$
with coefficients from the corresponding binomial distribution,
\begin{equation} \label{eq::canonical-def}
	F(p) = \sum_{n=0}^N {N \choose n}p^n(1-p)^{N-n}F(n),
\end{equation}
where $N$ is the number of sites in the system.
This, however, poses a subtle problem: our simulations are performed
for en ensemble of triangles in which $N$ may take on many values for
fixed $L$, and its mean value, $S(L)$, is non-integer. We solve this by replacing the binomial distribution with its
normal approximation, $\mathcal{N}\left(S(L)p,S(L)p(1-p)\right)$. This leads to
\begin{equation} \label{eq::canonical}
	R_L(p) = \frac{1}{\sigma \sqrt{2\pi} }\sum_{n=0}^{\lfloor S(L) \rfloor}
	\exp\left({\frac{-(n-\mu)^2}{2\sigma^2}}\right)R(n; L),
\end{equation}
with $\mu = S(L)p$, and $\sigma^2 = S(L)p(1-p)$.

\begin{figure}
	\includegraphics[width=0.9\columnwidth]{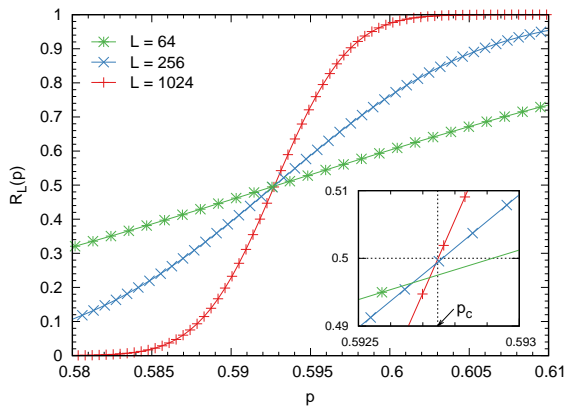}
	\caption{\label{fig:distr} The probability $R_L(p)$ that a triangle
		of side $L$ with occupation probability $p$ of its internal sites
		contains a three-leg percolating cluster.
		Inset: close-up of the intersection region.
	}
\end{figure}

It is expected \cite{Koza2019} that $R_L(p_\mathrm{c}) \to 1/2$ as
$L\to \infty$ (Fig.~\ref{fig:distr}).
We can use this property to define an $L$-dependent approximation
of the percolation threshold, $p^*_L$, defined as the solution to
\begin{equation} \label{eq:p*}
R_L(p^*_L) = 1/2.
\end{equation}
It was suggested \cite{Oliveira2003} that $p^*_L$ can be approximated with
\begin{equation} \label{eq::expansion-in-L}
	p^*_L \approx p_\mathrm{c} + L^{-1/\nu}  \sum_{k=0}^M A_k L^{-k},
\end{equation}
where $\nu = 4/3$ is a critical exponent \cite{Stauffer1994},
$A_k$ are some nonuniversal parameters, and $M$ is a cut-off.
However, in our case the sites forming "edge groups" are
situated inside the triangle, so the effective side length
may differ from $L$. We therefore introduce another parameter,
$\lambda$, that accounts for this uncertainty and helps to correct
for the truncation of higher-order terms
\cite{Levinshteln1975,Ziff92_PRL},
\begin{equation} \label{eq::expansion-in-L-lambda}
	p^*_L \approx p_\mathrm{c} + (L+\lambda)^{-1/\nu}  \sum_{k=0}^M A_k (L+\lambda)^{-k},\quad  L\ge L_\mathrm{min}.
\end{equation}
where $L_\mathrm{min}$ is the cut-off for the system size below which this
approximation is invalid.

The uncertainties of $p^*_L$ are estimated using
the bootstrap method \cite{nrc}.
Next, the Levenberg–Marquardt algorithm for nonlinear least squares
curve-fitting is applied to estimate the values of
$p_\mathrm{c}$, $\lambda$, and $A_k$
in Eq.~(\ref{eq::expansion-in-L-lambda}).
The cut-offs $L_\mathrm{min}$ and $M$ are determined from the requirement
that they should minimize the error estimate for $p_\mathrm{c}$.
Quality of the fit was monitored with the regression standard error
$s = \sqrt{\chi^2/\mathrm{dof}}$
(the square root of the chi-squared statistic per degree of freedom).
For a good fit, $s$ is close to or smaller than 1.

\section{Results\label{sec::Results}}

We performed simulations of nearest-neighbor site percolation
on a square lattice for triangles of side $L$
ranging from 4 to 1440 (all lengths in lattice units).
The total number of active sites inside the triangles was
$\approx 2\cdot 10^{14}$ for each $L\ge 16$, and  between $10^{12}$
and $10^{13}$ for each $L < 16$.
The uncertainties of $p^*_L$ were below $3 \cdot 10^{-7}$ for $L\ge 16$ and below $10^{-6}$ for $L<16$.

The first question we investigated was the convergence rate in
Eq.~(\ref{eq::expansion-in-L-lambda}). Since $R_L(p_\mathrm{c}) \to 1/2$
as $L\to\infty$, and we use this limiting value in Eq.~(\ref{eq:p*}),
$p^*_L - p_\mathrm{c}$ is expected to scale as $L^{-1 - 1/\nu}$ \cite{Ziff92_PRL,Oliveira2003},
which is equivalent to $A_0 = 0$ in (\ref{eq::expansion-in-L-lambda}).
When we set $M = 1$ in (\ref{eq::expansion-in-L-lambda}) and used
the best known value of $p_\mathrm{c}$, Eq.~(\ref{eq::pc_Jacobsen}),
we obtained  $A_0 =  3.5\cdot 10^{-6} \pm 2\cdot 10^{-6}$
with the regression standard error $s \approx 0.6$, in agreement with
the convergence rate $\sim L^{-1 - 1/\nu}$ = $L^{-7/4}$.

\begin{figure}
	\includegraphics[width=0.9\columnwidth]{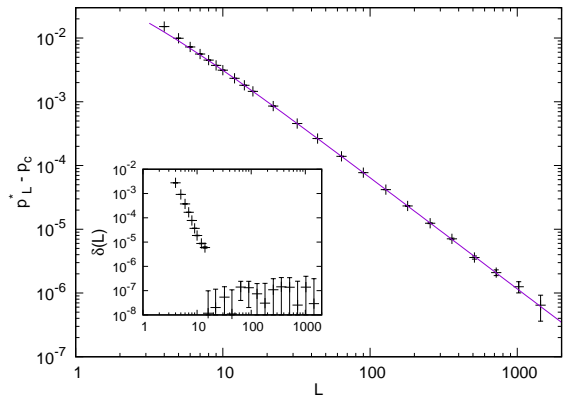}
	\caption{\label{fig:3}
		$p^*_L - p_\mathrm{c}$ (symbols) and its approximation
		$A_0/(L+\lambda)^{1+1/\nu}$ (line).
		Inset: $\delta(L) = |p^*_L - p_\mathrm{c} - A_0/(L+\lambda)^{1+1/\nu}|$,
		the error introduced by using (\ref{eq::expansion-in-L-lambda})
		with $M=1$ and $L_\mathrm{min}=16$.
	}
\end{figure}

Setting $A_0 = 0$, we obtained
$A_1 = 0.20707(6)$, $\lambda = 0.9942(25)$, and
\begin{equation} \label{eq::pc}
  p_\mathrm{c} =  0.592\,746\,10(4),
\end{equation}
with $L_\mathrm{min} = 16$ and the regression standard error
$s \approx 0.7$. This fit is shown in Fig.~\ref{fig:3}.
The value of $p_\mathrm{c}$ is in agreement with Eq.~(\ref{eq::pc_Jacobsen}).
In terms of measurement precision, our method turns out to be
at least on par with alternative simulation methods based
on Monte Carlo sampling (\cite{Feng2008,Lee2008} and references therein).
The value of $s$ being less than 1 indicates that the fit is
good: actually, the difference between 	$p^*_L$ and its approximation
$p_\mathrm{c} + A_1/(L+\lambda)^{1+1/\nu}$ is less than
$2\cdot 10^{-7}$ for all $L \ge 16$ (Fig.~\ref{fig:3}, inset).
For $4 \le L < 16$
it behaves roughly as $L^{-\omega}$ with $\omega = 6.0(4)$.
This suggests that $A_2 ,  A_3, A_4 \approx 0$ in
(\ref{eq::expansion-in-L-lambda}).
However, fits in this region are poor and the hypothesis that three
consecutive terms in (\ref{eq::expansion-in-L-lambda})
vanish should be taken with caution.

The role of $\lambda$ in (\ref{eq::expansion-in-L-lambda}) is to help
reduce both $M$ and $L_\mathrm{min}$. One could do without it
and set $\lambda = 0$, effectively reducing
(\ref{eq::expansion-in-L-lambda}) to (\ref{eq::expansion-in-L}).
This would require taking $M = 3$, with the total of 4 fitting
parameters instead of 3. The uncertainty of $p_\mathrm{c}$ thus
obtained would be higher by about 25\% compared to the
value reported in (\ref{eq::pc}), which is acceptable,
but the uncertainty of $A_1$ would be tripled.

\begin{figure}
	\includegraphics[width=0.9\columnwidth]{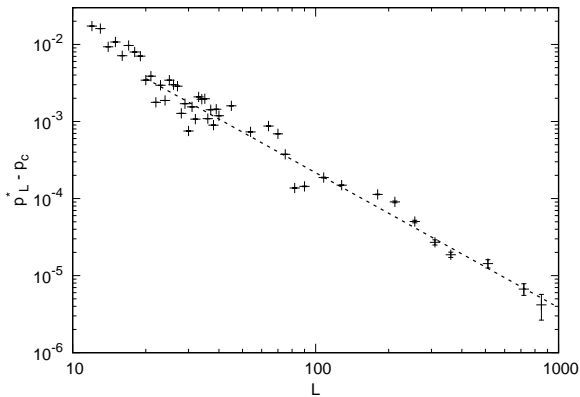}
	\caption{\label{fig:4}
		$p^*_L - p_\mathrm{c}$ for triangles with vertex $A$ fixed at $(0.5,0.5)$
		and angle $\alpha = 0$. The dashed line is a guide to the eye with the
		slope $-1-1/\nu = -1.75$.
	}
\end{figure}

Is randomization of the position and orientation of triangles necessary?
When we fixed vertex $A$ at (x, y) and $B$ at (x + L, y) with $x=y=0.5$,
we found that the left hand side of (\ref{eq::expansion-in-L-lambda})
still exhibits the $\sim L^{-1-1/\nu}$ behavior, but it also contains
a significant contribution of what can be regarded as noise
(Fig.~\ref{fig:4}).
Its magnitude is so large that it makes the method practically useless.
When we set $x=0.5$ and treated $y$ as a random
variable uniformly distributed in $(0,1)$, this ``noise'' was still present,
though its magnitude was much smaller (data not shown). Thus,
full randomization ot triangles' position and orientation
appears necessary for lattice-based systems.  However,
for continuous systems (e.g. percolation of
overlapping discs  or squares~\cite{Mertens2012})
this step can be omitted.

Using the canonical ensemble method to estimate $R_L(p)$ is crucial,
because it  works well even for small system sizes,
where the discrete nature of $R(n; L)$ is quite problematic for
other methods, e.g.\ those based on approximation.
Further studies are necessary to check whether applying
the exact canonical weighting could improve the validity
of Eq.~(\ref{eq::expansion-in-L-lambda}) for small $L$.

The simulations were run in parallel on several computers with varying 
computational power. 
Simulations depicted in Fig.~\ref{fig:3} would have taken about 1.5 years 
if done only on the PC on which this paper was written 
(a four-core processor running 8 threads in parallel at 3.4 GHz). 
The computational overhead introduced by randomization 
of the triangle orientation is about 100\% for very small systems ($L = 4$),
but quickly drops to $\approx 10\%$ for $L \gtrsim 50$. 
The highest (multi-threaded) computational efficiency is obtained for $L\approx 64$,  
for which one occupied site is processed at $\approx 33$ 
processor clock cycles. It decreases $\approx 5$ times for the smallest 
and largest system sizes considered here, the former due to 
the randomization overhead, the latter due to processor cache misses.

\section{Discussion\label{sec::Duscussion}}

The method presented here can be applied to other system shapes (with an arbitrary
division of their boundary into three compact regions), lattices, and 
cluster connectedness definitions, all with the universal three-leg crossing 
probability $1/2$. In contrast to this, universal spanning ("two-leg") and 
wrapping probabilities are known only for some particular system shapes, 
e.g.\ rectangles, and for more complicated cases would have to be treated as 
an additional unknown. A disadvantage of our method is a rather slow convergence
rate to the thermodynamic limit, most likely related to open 
boundary conditions~\cite{Hovi1996}.

The method can be generalized to more complex shapes, e.g., convex polygons.
The bounding box can be determined from the maximum and minimum values of
the vertex coordinates. Then one can iterate over each of the sites 
lying inside the bounding box and test
if it lies inside the polygon and if so, if it is adjacent
to any of the polygon's sides. This general method can be used for complex 
networks, e.g., Penrose tiling~\cite{Yonezawa1989I}.
A more efficient way is to shoot a small number of ``rays`` 
in such a way that they cut all internal sites of the polygon.
In the case depicted in Fig.~\ref{fig:geometry}
they could be shot vertically or horizontally along the grid lines. 
If a ray hits a convex polygon, it enters the system at some point $q_1$ and leaves it 
at some $q_2$ ($q_1$ and $q_2$  can coincide). 
Any lattice site between  $q_1$ and $q_2$ must be an internal site and all edge sites 
must lie in their close proximity. 
This reduces the computational complexity of discovering the edge sites 
from quadratic to linear in the characteristic system size. 
This method should be applicable for
all networks based on regular lattices, including networks 
with voids and bottlenecks~\cite{Haji-Akbari2009} and frustrated 
lattices~\cite{Haji-Akbari2015}. 
Once the geometry of the system has been established, the remaining steps 
are essentially the same as in traditional methods.

The angle by which each system is rotated can be arbitrary. 
Here we reduced it to the range from 0 o 15 degrees only because 
for equilateral triangles some code simplifications are then possible. 
For example, no sites adjacent to edge $AB$ in Fig.~\ref{fig:geometry} can have 
the same vertical coordinates.

Bond percolation can be treated in a very similar way. 
Rather than with active sites, one would deal with active bonds defined 
as those with at least one end lying inside the system. 
Edge bonds would be then defined as those that cut a given system edge.

Monte Carlo methods, including the one presented here, 
for simple planar lattice models 
have recently become overshadowed by transfer-matrix methods \cite{Jacobsen2015}. 
Still, we hope that our work can prove useful in more complicated cases, 
like networks with voids and bottlenecks, continuous models, 
or in higher-dimensions  \cite{Gori2015}.

\section{Conclusions\label{sec::Conclusions}}

We have shown that Monte-Carlo simulations in systems with
incompatible symmetries of their geometry and the underlying lattice
can be efficient in determining the percolation threshold.
The key step is randomization of the system orientation
and position relative to the lattice.
The computational overhead related to this additional step is acceptable.
Although the convergence rate to the thermodynamic limit
is slower than in some methods based on wrapping,
this is compensated by very small (or perhaps even vanishing)
values of several higher-order terms and the possibility of using
small lattices.

Three-leg clusters have proved useful in determining
the value of the percolation threshold.
Their advantage is that the universal crossing probability
associated with them is geometry-independent, which opens the room
for further improvements of the method.


\begin{thebibliography}{44}%
	\makeatletter
	\providecommand \@ifxundefined [1]{%
		\@ifx{#1\undefined}
	}%
	\providecommand \@ifnum [1]{%
		\ifnum #1\expandafter \@firstoftwo
		\else \expandafter \@secondoftwo
		\fi
	}%
	\providecommand \@ifx [1]{%
		\ifx #1\expandafter \@firstoftwo
		\else \expandafter \@secondoftwo
		\fi
	}%
	\providecommand \natexlab [1]{#1}%
	\providecommand \enquote  [1]{``#1''}%
	\providecommand \bibnamefont  [1]{#1}%
	\providecommand \bibfnamefont [1]{#1}%
	\providecommand \citenamefont [1]{#1}%
	\providecommand \href@noop [0]{\@secondoftwo}%
	\providecommand \href [0]{\begingroup \@sanitize@url \@href}%
	\providecommand \@href[1]{\@@startlink{#1}\@@href}%
	\providecommand \@@href[1]{\endgroup#1\@@endlink}%
	\providecommand \@sanitize@url [0]{\catcode `\\12\catcode `\$12\catcode
		`\&12\catcode `\#12\catcode `\^12\catcode `\_12\catcode `\%12\relax}%
	\providecommand \@@startlink[1]{}%
	\providecommand \@@endlink[0]{}%
	\providecommand \url  [0]{\begingroup\@sanitize@url \@url }%
	\providecommand \@url [1]{\endgroup\@href {#1}{\urlprefix }}%
	\providecommand \urlprefix  [0]{URL }%
	\providecommand \Eprint [0]{\href }%
	\providecommand \doibase [0]{https://doi.org/}%
	\providecommand \selectlanguage [0]{\@gobble}%
	\providecommand \bibinfo  [0]{\@secondoftwo}%
	\providecommand \bibfield  [0]{\@secondoftwo}%
	\providecommand \translation [1]{[#1]}%
	\providecommand \BibitemOpen [0]{}%
	\providecommand \bibitemStop [0]{}%
	\providecommand \bibitemNoStop [0]{.\EOS\space}%
	\providecommand \EOS [0]{\spacefactor3000\relax}%
	\providecommand \BibitemShut  [1]{\csname bibitem#1\endcsname}%
	\let\auto@bib@innerbib\@empty
	\bibitem [{\citenamefont {Hunt}(2001)}]{Hunt2001}%
	\BibitemOpen
	\bibfield  {author} {\bibinfo {author} {\bibfnamefont {A.}~\bibnamefont
			{Hunt}},\ }\bibfield  {title} {\bibinfo {title} {Applications of percolation
			theory to porous media with distributed local conductances},\ }\href
	{https://doi.org/https://doi.org/10.1016/S0309-1708(00)00058-0} {\bibfield
		{journal} {\bibinfo  {journal} {Advances in Water Resources}\ }\textbf
		{\bibinfo {volume} {24}},\ \bibinfo {pages} {279} (\bibinfo {year}
		{2001})}\BibitemShut {NoStop}%
	\bibitem [{\citenamefont {Bessaguet}\ \emph {et~al.}(2019)\citenamefont
		{Bessaguet}, \citenamefont {Dantras}, \citenamefont {Michon}, \citenamefont
		{Chevalier}, \citenamefont {Laffont},\ and\ \citenamefont
		{Lacabanne}}]{Besseaguet2019}%
	\BibitemOpen
	\bibfield  {author} {\bibinfo {author} {\bibfnamefont {C.}~\bibnamefont
			{Bessaguet}}, \bibinfo {author} {\bibfnamefont {E.}~\bibnamefont {Dantras}},
		\bibinfo {author} {\bibfnamefont {G.}~\bibnamefont {Michon}}, \bibinfo
		{author} {\bibfnamefont {M.}~\bibnamefont {Chevalier}}, \bibinfo {author}
		{\bibfnamefont {L.}~\bibnamefont {Laffont}},\ and\ \bibinfo {author}
		{\bibfnamefont {C.}~\bibnamefont {Lacabanne}},\ }\bibfield  {title} {\bibinfo
		{title} {Electrical behavior of a graphene/{PEKK} and carbon black/{PEKK}
			nanocomposites in the vicinity of the percolation threshold},\ }\href
	{https://doi.org/https://doi.org/10.1016/j.jnoncrysol.2019.02.017} {\bibfield
		{journal} {\bibinfo  {journal} {Journal of Non-Crystalline Solids}\ }\textbf
		{\bibinfo {volume} {512}},\ \bibinfo {pages} {1} (\bibinfo {year}
		{2019})}\BibitemShut {NoStop}%
	\bibitem [{\citenamefont {Sahimi}(1993)}]{Sahimi1993}%
	\BibitemOpen
	\bibfield  {author} {\bibinfo {author} {\bibfnamefont {M.}~\bibnamefont
			{Sahimi}},\ }\bibfield  {title} {\bibinfo {title} {Flow phenomena in rocks:
			from continuum models to fractals, percolation, cellular automata, and
			simulated annealing},\ }\href {https://doi.org/10.1103/RevModPhys.65.1393}
	{\bibfield  {journal} {\bibinfo  {journal} {Rev. Mod. Phys.}\ }\textbf
		{\bibinfo {volume} {65}},\ \bibinfo {pages} {1393} (\bibinfo {year}
		{1993})}\BibitemShut {NoStop}%
	\bibitem [{\citenamefont {Bolandtaba}\ and\ \citenamefont
		{Skauge}(2011)}]{Bolandtaba2011}%
	\BibitemOpen
	\bibfield  {author} {\bibinfo {author} {\bibfnamefont {S.}~\bibnamefont
			{Bolandtaba}}\ and\ \bibinfo {author} {\bibfnamefont {A.}~\bibnamefont
			{Skauge}},\ }\bibfield  {title} {\bibinfo {title} {Network modeling of {EOR}
			processes: a combined invasion percolation and dynamic model for mobilization
			of trapped oil},\ }\href@noop {} {\bibfield  {journal} {\bibinfo  {journal}
			{Transport in porous media}\ }\textbf {\bibinfo {volume} {89}},\ \bibinfo
		{pages} {357} (\bibinfo {year} {2011})}\BibitemShut {NoStop}%
	\bibitem [{\citenamefont {Ziff}(2021)}]{Ziff2021}%
	\BibitemOpen
	\bibfield  {author} {\bibinfo {author} {\bibfnamefont {R.~M.}\ \bibnamefont
			{Ziff}},\ }\bibfield  {title} {\bibinfo {title} {Percolation and the
			pandemic},\ }\href
	{https://doi.org/https://doi.org/10.1016/j.physa.2020.125723} {\bibfield
		{journal} {\bibinfo  {journal} {Physica A: Statistical Mechanics and its
				Applications}\ }\textbf {\bibinfo {volume} {568}},\ \bibinfo {pages} {125723}
		(\bibinfo {year} {2021})}\BibitemShut {NoStop}%
	\bibitem [{\citenamefont {Shtein}\ \emph {et~al.}(2015)\citenamefont {Shtein},
		\citenamefont {Nadiv}, \citenamefont {Buzaglo}, \citenamefont {Kahil},\ and\
		\citenamefont {Regev}}]{Shtein2015}%
	\BibitemOpen
	\bibfield  {author} {\bibinfo {author} {\bibfnamefont {M.}~\bibnamefont
			{Shtein}}, \bibinfo {author} {\bibfnamefont {R.}~\bibnamefont {Nadiv}},
		\bibinfo {author} {\bibfnamefont {M.}~\bibnamefont {Buzaglo}}, \bibinfo
		{author} {\bibfnamefont {K.}~\bibnamefont {Kahil}},\ and\ \bibinfo {author}
		{\bibfnamefont {O.}~\bibnamefont {Regev}},\ }\bibfield  {title} {\bibinfo
		{title} {Thermally conductive graphene-polymer composites: size, percolation,
			and synergy effects},\ }\href@noop {} {\bibfield  {journal} {\bibinfo
			{journal} {Chemistry of Materials}\ }\textbf {\bibinfo {volume} {27}},\
		\bibinfo {pages} {2100} (\bibinfo {year} {2015})}\BibitemShut {NoStop}%
	\bibitem [{\citenamefont {Stauffer}\ and\ \citenamefont
		{Aharony}(1994)}]{Stauffer1994}%
	\BibitemOpen
	\bibfield  {author} {\bibinfo {author} {\bibfnamefont {D.}~\bibnamefont
			{Stauffer}}\ and\ \bibinfo {author} {\bibfnamefont {A.}~\bibnamefont
			{Aharony}},\ }\href@noop {} {\emph {\bibinfo {title} {Introduction to
				Percolation Theory}}},\ \bibinfo {edition} {2nd}\ ed.\ (\bibinfo  {publisher}
	{Taylor and Francis},\ \bibinfo {address} {London},\ \bibinfo {year}
	{1994})\BibitemShut {NoStop}%
	\bibitem [{\citenamefont {Cardy}(1992)}]{Cardy1992}%
	\BibitemOpen
	\bibfield  {author} {\bibinfo {author} {\bibfnamefont {J.~L.}\ \bibnamefont
			{Cardy}},\ }\bibfield  {title} {\bibinfo {title} {Critical percolation in
			finite geometries},\ }\href {https://doi.org/10.1088/0305-4470/25/4/009}
	{\bibfield  {journal} {\bibinfo  {journal} {Journal of Physics A:
				Mathematical and General}\ }\textbf {\bibinfo {volume} {25}},\ \bibinfo
		{pages} {L201} (\bibinfo {year} {1992})}\BibitemShut {NoStop}%
	\bibitem [{\citenamefont {Lawler}\ \emph {et~al.}(2002)\citenamefont {Lawler},
		\citenamefont {Schramm},\ and\ \citenamefont {Werner}}]{Lawler2001}%
	\BibitemOpen
	\bibfield  {author} {\bibinfo {author} {\bibfnamefont {G.}~\bibnamefont
			{Lawler}}, \bibinfo {author} {\bibfnamefont {O.}~\bibnamefont {Schramm}},\
		and\ \bibinfo {author} {\bibfnamefont {W.}~\bibnamefont {Werner}},\
	}\bibfield  {title} {\bibinfo {title} {One-arm exponent for critical {2D}
			percolation},\ }\href
	{http://www.math.washington.edu/~ejpecp/EjpVol7/paper2.abs.html} {\bibfield
		{journal} {\bibinfo  {journal} {Electronic Journal of Probability}\ }\textbf
		{\bibinfo {volume} {7}},\ \bibinfo {pages} {1} (\bibinfo {year}
		{2002})}\BibitemShut {NoStop}%
	\bibitem [{\citenamefont {Smirnov}(2001)}]{Smirnov2001ci}%
	\BibitemOpen
	\bibfield  {author} {\bibinfo {author} {\bibfnamefont {S.}~\bibnamefont
			{Smirnov}},\ }\bibfield  {title} {\bibinfo {title} {Critical percolation in
			the plane: conformal invariance, cardy's formula, scaling limits},\ }\href
	{https://doi.org/https://doi.org/10.1016/S0764-4442(01)01991-7} {\bibfield
		{journal} {\bibinfo  {journal} {Comptes Rendus de l'Académie des Sciences -
				Series I - Mathematics}\ }\textbf {\bibinfo {volume} {333}},\ \bibinfo
		{pages} {239 } (\bibinfo {year} {2001})}\BibitemShut {NoStop}%
	\bibitem [{\citenamefont {Flores}\ \emph {et~al.}(2017)\citenamefont {Flores},
		\citenamefont {Simmons}, \citenamefont {Kleban},\ and\ \citenamefont
		{Ziff}}]{Flores2017}%
	\BibitemOpen
	\bibfield  {author} {\bibinfo {author} {\bibfnamefont {S.~M.}\ \bibnamefont
			{Flores}}, \bibinfo {author} {\bibfnamefont {J.~J.~H.}\ \bibnamefont
			{Simmons}}, \bibinfo {author} {\bibfnamefont {P.}~\bibnamefont {Kleban}},\
		and\ \bibinfo {author} {\bibfnamefont {R.~M.}\ \bibnamefont {Ziff}},\
	}\bibfield  {title} {\bibinfo {title} {A formula for crossing probabilities
			of critical systems inside polygons},\ }\href
	{https://doi.org/10.1088/1751-8121/50/6/064005} {\bibfield  {journal}
		{\bibinfo  {journal} {Journal of Physics A: Mathematical and Theoretical}\
		}\textbf {\bibinfo {volume} {50}},\ \bibinfo {pages} {064005} (\bibinfo
		{year} {2017})}\BibitemShut {NoStop}%
	\bibitem [{\citenamefont {Sykes}\ and\ \citenamefont
		{Essam}(1964)}]{Sykes1964}%
	\BibitemOpen
	\bibfield  {author} {\bibinfo {author} {\bibfnamefont {M.~F.}\ \bibnamefont
			{Sykes}}\ and\ \bibinfo {author} {\bibfnamefont {J.~W.}\ \bibnamefont
			{Essam}},\ }\bibfield  {title} {\bibinfo {title} {Exact critical percolation
			probabilities for site and bond problems in two dimensions},\ }\href
	{https://doi.org/10.1063/1.1704215} {\bibfield  {journal} {\bibinfo
			{journal} {Journal of Mathematical Physics}\ }\textbf {\bibinfo {volume}
			{5}},\ \bibinfo {pages} {1117} (\bibinfo {year} {1964})},\ \Eprint
	{https://arxiv.org/abs/https://doi.org/10.1063/1.1704215}
	{https://doi.org/10.1063/1.1704215} \BibitemShut {NoStop}%
	\bibitem [{\citenamefont {Kesten}(1980)}]{Kesten1980}%
	\BibitemOpen
	\bibfield  {author} {\bibinfo {author} {\bibfnamefont {H.}~\bibnamefont
			{Kesten}},\ }\bibfield  {title} {\bibinfo {title} {The critical probability
			of bond percolation on the square lattice equals 1/2},\ }\href
	{https://doi.org/10.1007/BF01197577} {\bibfield  {journal} {\bibinfo
			{journal} {Communications in Mathematical Physics}\ }\textbf {\bibinfo
			{volume} {74}},\ \bibinfo {pages} {41} (\bibinfo {year} {1980})}\BibitemShut
	{NoStop}%
	\bibitem [{\citenamefont {Wierman}(2009)}]{Wierman2009}%
	\BibitemOpen
	\bibfield  {author} {\bibinfo {author} {\bibfnamefont {J.~C.}\ \bibnamefont
			{Wierman}},\ }\bibinfo {title} {Percolation thresholds, exact},\ in\ \href
	{https://doi.org/10.1007/978-0-387-30440-3_390} {\emph {\bibinfo {booktitle}
			{Encyclopedia of Complexity and Systems Science}}},\ \bibinfo {editor}
	{edited by\ \bibinfo {editor} {\bibfnamefont {R.~A.}\ \bibnamefont
			{Meyers}}}\ (\bibinfo  {publisher} {Springer New York},\ \bibinfo {address}
	{New York, NY},\ \bibinfo {year} {2009})\ pp.\ \bibinfo {pages}
	{6579--6587}\BibitemShut {NoStop}%
	\bibitem [{\citenamefont {Leath}(1976)}]{Leath1976}%
	\BibitemOpen
	\bibfield  {author} {\bibinfo {author} {\bibfnamefont {P.~L.}\ \bibnamefont
			{Leath}},\ }\bibfield  {title} {\bibinfo {title} {Cluster size and boundary
			distribution near percolation threshold},\ }\href
	{https://doi.org/10.1103/PhysRevB.14.5046} {\bibfield  {journal} {\bibinfo
			{journal} {Phys. Rev. B}\ }\textbf {\bibinfo {volume} {14}},\ \bibinfo
		{pages} {5046} (\bibinfo {year} {1976})}\BibitemShut {NoStop}%
	\bibitem [{\citenamefont {Lorenz}\ and\ \citenamefont
		{Ziff}(1998)}]{Lorenz1998}%
	\BibitemOpen
	\bibfield  {author} {\bibinfo {author} {\bibfnamefont {C.~D.}\ \bibnamefont
			{Lorenz}}\ and\ \bibinfo {author} {\bibfnamefont {R.~M.}\ \bibnamefont
			{Ziff}},\ }\bibfield  {title} {\bibinfo {title} {Precise determination of the
			bond percolation thresholds and finite-size scaling corrections for the sc,
			fcc, and bcc lattices},\ }\href {https://doi.org/10.1103/PhysRevE.57.230}
	{\bibfield  {journal} {\bibinfo  {journal} {Phys. Rev. E}\ }\textbf {\bibinfo
			{volume} {57}},\ \bibinfo {pages} {230} (\bibinfo {year} {1998})}\BibitemShut
	{NoStop}%
	\bibitem [{\citenamefont {Xun}\ \emph {et~al.}(2021)\citenamefont {Xun},
		\citenamefont {Hao},\ and\ \citenamefont {Ziff}}]{Xun2021}%
	\BibitemOpen
	\bibfield  {author} {\bibinfo {author} {\bibfnamefont {Z.}~\bibnamefont
			{Xun}}, \bibinfo {author} {\bibfnamefont {D.}~\bibnamefont {Hao}},\ and\
		\bibinfo {author} {\bibfnamefont {R.~M.}\ \bibnamefont {Ziff}},\ }\bibfield
	{title} {\bibinfo {title} {Site percolation on square and simple cubic
			lattices with extended neighborhoods and their continuum limit},\ }\href
	{https://doi.org/10.1103/PhysRevE.103.022126} {\bibfield  {journal} {\bibinfo
			{journal} {Phys. Rev. E}\ }\textbf {\bibinfo {volume} {103}},\ \bibinfo
		{pages} {022126} (\bibinfo {year} {2021})}\BibitemShut {NoStop}%
	\bibitem [{\citenamefont {Voss}(1984)}]{Voss1984}%
	\BibitemOpen
	\bibfield  {author} {\bibinfo {author} {\bibfnamefont {R.~F.}\ \bibnamefont
			{Voss}},\ }\bibfield  {title} {\bibinfo {title} {The fractal dimension of
			percolation cluster hulls},\ }\href
	{https://doi.org/10.1088/0305-4470/17/7/001} {\bibfield  {journal} {\bibinfo
			{journal} {Journal of Physics A: Mathematical and General}\ }\textbf
		{\bibinfo {volume} {17}},\ \bibinfo {pages} {L373} (\bibinfo {year}
		{1984})}\BibitemShut {NoStop}%
	\bibitem [{\citenamefont {Ziff}\ \emph {et~al.}(1984)\citenamefont {Ziff},
		\citenamefont {Cummings},\ and\ \citenamefont {Stells}}]{Ziff1984}%
	\BibitemOpen
	\bibfield  {author} {\bibinfo {author} {\bibfnamefont {R.~M.}\ \bibnamefont
			{Ziff}}, \bibinfo {author} {\bibfnamefont {P.~T.}\ \bibnamefont {Cummings}},\
		and\ \bibinfo {author} {\bibfnamefont {G.}~\bibnamefont {Stells}},\
	}\bibfield  {title} {\bibinfo {title} {Generation of percolation cluster
			perimeters by a random walk},\ }\href
	{https://doi.org/10.1088/0305-4470/17/15/018} {\bibfield  {journal} {\bibinfo
			{journal} {Journal of Physics A: Mathematical and General}\ }\textbf
		{\bibinfo {volume} {17}},\ \bibinfo {pages} {3009} (\bibinfo {year}
		{1984})}\BibitemShut {NoStop}%
	\bibitem [{\citenamefont {Rosso}\ \emph {et~al.}(1985)\citenamefont {Rosso},
		\citenamefont {Gouyet},\ and\ \citenamefont {Sapoval}}]{Rosso1985}%
	\BibitemOpen
	\bibfield  {author} {\bibinfo {author} {\bibfnamefont {M.}~\bibnamefont
			{Rosso}}, \bibinfo {author} {\bibfnamefont {J.~F.}\ \bibnamefont {Gouyet}},\
		and\ \bibinfo {author} {\bibfnamefont {B.}~\bibnamefont {Sapoval}},\
	}\bibfield  {title} {\bibinfo {title} {Determination of percolation
			probability from the use of a concentration gradient},\ }\href
	{https://doi.org/10.1103/PhysRevB.32.6053} {\bibfield  {journal} {\bibinfo
			{journal} {Phys. Rev. B}\ }\textbf {\bibinfo {volume} {32}},\ \bibinfo
		{pages} {6053} (\bibinfo {year} {1985})}\BibitemShut {NoStop}%
	\bibitem [{\citenamefont {Ziff}\ and\ \citenamefont
		{Sapoval}(1986)}]{Ziff1986}%
	\BibitemOpen
	\bibfield  {author} {\bibinfo {author} {\bibfnamefont {R.~M.}\ \bibnamefont
			{Ziff}}\ and\ \bibinfo {author} {\bibfnamefont {B.}~\bibnamefont {Sapoval}},\
	}\bibfield  {title} {\bibinfo {title} {The efficient determination of the
			percolation threshold by a frontier-generating walk in a gradient},\ }\href
	{https://doi.org/10.1088/0305-4470/19/18/010} {\bibfield  {journal} {\bibinfo
			{journal} {Journal of Physics A: Mathematical and General}\ }\textbf
		{\bibinfo {volume} {19}},\ \bibinfo {pages} {L1169} (\bibinfo {year}
		{1986})}\BibitemShut {NoStop}%
	\bibitem [{\citenamefont {Tencer}\ and\ \citenamefont
		{Forsberg}(2021)}]{Tencer2021}%
	\BibitemOpen
	\bibfield  {author} {\bibinfo {author} {\bibfnamefont {J.}~\bibnamefont
			{Tencer}}\ and\ \bibinfo {author} {\bibfnamefont {K.~M.}\ \bibnamefont
			{Forsberg}},\ }\bibfield  {title} {\bibinfo {title} {Postprocessing
			techniques for gradient percolation predictions on the square lattice},\
	}\href {https://doi.org/10.1103/PhysRevE.103.012115} {\bibfield  {journal}
		{\bibinfo  {journal} {Phys. Rev. E}\ }\textbf {\bibinfo {volume} {103}},\
		\bibinfo {pages} {012115} (\bibinfo {year} {2021})}\BibitemShut {NoStop}%
	\bibitem [{\citenamefont {Newman}\ and\ \citenamefont
		{Ziff}(2001)}]{Newman2001}%
	\BibitemOpen
	\bibfield  {author} {\bibinfo {author} {\bibfnamefont {M.~E.~J.}\
			\bibnamefont {Newman}}\ and\ \bibinfo {author} {\bibfnamefont {R.~M.}\
			\bibnamefont {Ziff}},\ }\bibfield  {title} {\bibinfo {title} {Fast {M}onte
			{C}arlo algorithm for site or bond percolation},\ }\href
	{https://doi.org/10.1103/PhysRevE.64.016706} {\bibfield  {journal} {\bibinfo
			{journal} {Phys. Rev. E}\ }\textbf {\bibinfo {volume} {64}},\ \bibinfo
		{pages} {016706} (\bibinfo {year} {2001})}\BibitemShut {NoStop}%
	\bibitem [{\citenamefont {Wang}\ \emph {et~al.}(2013)\citenamefont {Wang},
		\citenamefont {Zhou}, \citenamefont {Zhang}, \citenamefont {Garoni},\ and\
		\citenamefont {Deng}}]{Wang2013}%
	\BibitemOpen
	\bibfield  {author} {\bibinfo {author} {\bibfnamefont {J.}~\bibnamefont
			{Wang}}, \bibinfo {author} {\bibfnamefont {Z.}~\bibnamefont {Zhou}}, \bibinfo
		{author} {\bibfnamefont {W.}~\bibnamefont {Zhang}}, \bibinfo {author}
		{\bibfnamefont {T.~M.}\ \bibnamefont {Garoni}},\ and\ \bibinfo {author}
		{\bibfnamefont {Y.}~\bibnamefont {Deng}},\ }\bibfield  {title} {\bibinfo
		{title} {Bond and site percolation in three dimensions},\ }\href
	{https://doi.org/10.1103/PhysRevE.87.052107} {\bibfield  {journal} {\bibinfo
			{journal} {Phys. Rev. E}\ }\textbf {\bibinfo {volume} {87}},\ \bibinfo
		{pages} {052107} (\bibinfo {year} {2013})}\BibitemShut {NoStop}%
	\bibitem [{\citenamefont {Koza}\ and\ \citenamefont
		{Po{\l}a}(2016)}]{Koza2016}%
	\BibitemOpen
	\bibfield  {author} {\bibinfo {author} {\bibfnamefont {Z.}~\bibnamefont
			{Koza}}\ and\ \bibinfo {author} {\bibfnamefont {J.}~\bibnamefont {Po{\l}a}},\
	}\bibfield  {title} {\bibinfo {title} {From discrete to continuous
			percolation in dimensions 3 to 7},\ }\href
	{https://doi.org/10.1088/1742-5468/2016/10/103206} {\bibfield  {journal}
		{\bibinfo  {journal} {Journal of Statistical Mechanics: Theory and
				Experiment}\ }\textbf {\bibinfo {volume} {2016}},\ \bibinfo {pages} {103206}
		(\bibinfo {year} {2016})}\BibitemShut {NoStop}%
	\bibitem [{\citenamefont {Ziff}(1992)}]{Ziff92_PRL}%
	\BibitemOpen
	\bibfield  {author} {\bibinfo {author} {\bibfnamefont {R.~M.}\ \bibnamefont
			{Ziff}},\ }\bibfield  {title} {\bibinfo {title} {Spanning probability in {2D}
			percolation},\ }\href {https://doi.org/10.1103/PhysRevLett.69.2670}
	{\bibfield  {journal} {\bibinfo  {journal} {Phys. Rev. Lett.}\ }\textbf
		{\bibinfo {volume} {69}},\ \bibinfo {pages} {2670} (\bibinfo {year}
		{1992})}\BibitemShut {NoStop}%
	\bibitem [{\citenamefont {de~Oliveira}\ \emph {et~al.}(2003)\citenamefont
		{de~Oliveira}, \citenamefont {N{\'o}rbrega},\ and\ \citenamefont
		{Stauffer}}]{Oliveira2003}%
	\BibitemOpen
	\bibfield  {author} {\bibinfo {author} {\bibfnamefont {P.~M.~C.}\
			\bibnamefont {de~Oliveira}}, \bibinfo {author} {\bibfnamefont {R.~A.}\
			\bibnamefont {N{\'o}rbrega}},\ and\ \bibinfo {author} {\bibfnamefont
			{D.}~\bibnamefont {Stauffer}},\ }\bibfield  {title} {\bibinfo {title}
		{Corrections to finite size scaling in percolation},\ }\href
	{https://doi.org/10.1590/S0103-97332003000300025} {\bibfield  {journal}
		{\bibinfo  {journal} {Brazilian Journal of Physics}\ }\textbf {\bibinfo
			{volume} {33}},\ \bibinfo {pages} {616 } (\bibinfo {year}
		{2003})}\BibitemShut {NoStop}%
	\bibitem [{\citenamefont {Torquato}\ and\ \citenamefont
		{Jiao}(2012)}]{Torquato12b}%
	\BibitemOpen
	\bibfield  {author} {\bibinfo {author} {\bibfnamefont {S.}~\bibnamefont
			{Torquato}}\ and\ \bibinfo {author} {\bibfnamefont {Y.}~\bibnamefont
			{Jiao}},\ }\bibfield  {title} {\bibinfo {title} {Effect of dimensionality on
			the continuum percolation of overlapping hyperspheres and hypercubes. {II}.
			{S}imulation results and analyses},\ }\href
	{https://doi.org/http://dx.doi.org/10.1063/1.4742750} {\bibfield  {journal}
		{\bibinfo  {journal} {The Journal of Chemical Physics}\ }\textbf {\bibinfo
			{volume} {137}},\ \bibinfo {eid} {074106} (\bibinfo {year}
		{2012})}\BibitemShut {NoStop}%
	\bibitem [{\citenamefont {Quintanilla}\ \emph {et~al.}(2000)\citenamefont
		{Quintanilla}, \citenamefont {Torquato},\ and\ \citenamefont
		{Ziff}}]{Quintanilla2000}%
	\BibitemOpen
	\bibfield  {author} {\bibinfo {author} {\bibfnamefont {J.}~\bibnamefont
			{Quintanilla}}, \bibinfo {author} {\bibfnamefont {S.}~\bibnamefont
			{Torquato}},\ and\ \bibinfo {author} {\bibfnamefont {R.~M.}\ \bibnamefont
			{Ziff}},\ }\bibfield  {title} {\bibinfo {title} {Efficient measurement of the
			percolation threshold for fully penetrable discs},\ }\href
	{https://doi.org/10.1088/0305-4470/33/42/104} {\bibfield  {journal} {\bibinfo
			{journal} {Journal of Physics A: Mathematical and General}\ }\textbf
		{\bibinfo {volume} {33}},\ \bibinfo {pages} {L399} (\bibinfo {year}
		{2000})}\BibitemShut {NoStop}%
	\bibitem [{\citenamefont {Pruessner}\ and\ \citenamefont
		{Moloney}(2003)}]{Pruessner03}%
	\BibitemOpen
	\bibfield  {author} {\bibinfo {author} {\bibfnamefont {G.}~\bibnamefont
			{Pruessner}}\ and\ \bibinfo {author} {\bibfnamefont {N.~R.}\ \bibnamefont
			{Moloney}},\ }\bibfield  {title} {\bibinfo {title} {Numerical results for
			crossing, spanning and wrapping in two-dimensional percolation},\ }\href
	{http://stacks.iop.org/0305-4470/36/i=44/a=003} {\bibfield  {journal}
		{\bibinfo  {journal} {Journal of Physics A: Mathematical and General}\
		}\textbf {\bibinfo {volume} {36}},\ \bibinfo {pages} {11213} (\bibinfo {year}
		{2003})}\BibitemShut {NoStop}%
	\bibitem [{\citenamefont {Yang}\ \emph {et~al.}(2013)\citenamefont {Yang},
		\citenamefont {Zhou},\ and\ \citenamefont {Li}}]{Yang2013}%
	\BibitemOpen
	\bibfield  {author} {\bibinfo {author} {\bibfnamefont {Y.}~\bibnamefont
			{Yang}}, \bibinfo {author} {\bibfnamefont {S.}~\bibnamefont {Zhou}},\ and\
		\bibinfo {author} {\bibfnamefont {Y.}~\bibnamefont {Li}},\ }\bibfield
	{title} {\bibinfo {title} {Square++: Making a connection game win-lose
			complementary and playing-fair},\ }\href
	{https://doi.org/https://doi.org/10.1016/j.entcom.2012.10.004} {\bibfield
		{journal} {\bibinfo  {journal} {Entertainment Computing}\ }\textbf {\bibinfo
			{volume} {4}},\ \bibinfo {pages} {105} (\bibinfo {year} {2013})}\BibitemShut
	{NoStop}%
	\bibitem [{\citenamefont {Feng}\ \emph {et~al.}(2008)\citenamefont {Feng},
		\citenamefont {Deng},\ and\ \citenamefont {Bl\"ote}}]{Feng2008}%
	\BibitemOpen
	\bibfield  {author} {\bibinfo {author} {\bibfnamefont {X.}~\bibnamefont
			{Feng}}, \bibinfo {author} {\bibfnamefont {Y.}~\bibnamefont {Deng}},\ and\
		\bibinfo {author} {\bibfnamefont {H.~W.~J.}\ \bibnamefont {Bl\"ote}},\
	}\bibfield  {title} {\bibinfo {title} {Percolation transitions in two
			dimensions},\ }\href {https://doi.org/10.1103/PhysRevE.78.031136} {\bibfield
		{journal} {\bibinfo  {journal} {Phys. Rev. E}\ }\textbf {\bibinfo {volume}
			{78}},\ \bibinfo {pages} {031136} (\bibinfo {year} {2008})}\BibitemShut
	{NoStop}%
	\bibitem [{\citenamefont {Jacobsen}(2015)}]{Jacobsen2015}%
	\BibitemOpen
	\bibfield  {author} {\bibinfo {author} {\bibfnamefont {J.~L.}\ \bibnamefont
			{Jacobsen}},\ }\bibfield  {title} {\bibinfo {title} {Critical points of
			{P}otts and {O}({N}) models from eigenvalue identities in periodic
			{T}emperley-{L}ieb algebras},\ }\href
	{http://stacks.iop.org/1751-8121/48/i=45/a=454003} {\bibfield  {journal}
		{\bibinfo  {journal} {Journal of Physics A: Mathematical and Theoretical}\
		}\textbf {\bibinfo {volume} {48}},\ \bibinfo {pages} {454003} (\bibinfo
		{year} {2015})}\BibitemShut {NoStop}%
	\bibitem [{\citenamefont {Koza}(2019)}]{Koza2019}%
	\BibitemOpen
	\bibfield  {author} {\bibinfo {author} {\bibfnamefont {Z.}~\bibnamefont
			{Koza}},\ }\bibfield  {title} {\bibinfo {title} {Critical $p=1/2$ in
			percolation on semi-infinite strips},\ }\href
	{https://doi.org/10.1103/PhysRevE.100.042115} {\bibfield  {journal} {\bibinfo
			{journal} {Phys. Rev. E}\ }\textbf {\bibinfo {volume} {100}},\ \bibinfo
		{pages} {042115} (\bibinfo {year} {2019})}\BibitemShut {NoStop}%
	\bibitem [{\citenamefont {Cormen}\ \emph {et~al.}(2009)\citenamefont {Cormen},
		\citenamefont {Leiserson}, \citenamefont {Rivest},\ and\ \citenamefont
		{Stein}}]{Cormen2009}%
	\BibitemOpen
	\bibfield  {author} {\bibinfo {author} {\bibfnamefont {T.~H.}\ \bibnamefont
			{Cormen}}, \bibinfo {author} {\bibfnamefont {C.~E.}\ \bibnamefont
			{Leiserson}}, \bibinfo {author} {\bibfnamefont {R.~L.}\ \bibnamefont
			{Rivest}},\ and\ \bibinfo {author} {\bibfnamefont {C.}~\bibnamefont
			{Stein}},\ }\href@noop {} {\emph {\bibinfo {title} {Introduction to
				algorithms}}}\ (\bibinfo  {publisher} {MIT press},\ \bibinfo {year}
	{2009})\BibitemShut {NoStop}%
	\bibitem [{\citenamefont {Levinshteln}\ and\ \citenamefont
		{Efros}(1975)}]{Levinshteln1975}%
	\BibitemOpen
	\bibfield  {author} {\bibinfo {author} {\bibfnamefont {M.}~\bibnamefont
			{Levinshteln}}\ and\ \bibinfo {author} {\bibfnamefont {L.}~\bibnamefont
			{Efros}},\ }\bibfield  {title} {\bibinfo {title} {The relation between the
			critical exponents of percolation theory},\ }\href@noop {} {\bibfield
		{journal} {\bibinfo  {journal} {Zh. Eksp. Teor. Fiz}\ }\textbf {\bibinfo
			{volume} {69}},\ \bibinfo {pages} {396} (\bibinfo {year} {1975})}\BibitemShut
	{NoStop}%
	\bibitem [{\citenamefont {Press}\ \emph {et~al.}(2007)\citenamefont {Press},
		\citenamefont {Teukolsky}, \citenamefont {Vetterling},\ and\ \citenamefont
		{Flannery}}]{nrc}%
	\BibitemOpen
	\bibfield  {author} {\bibinfo {author} {\bibfnamefont {W.~H.}\ \bibnamefont
			{Press}}, \bibinfo {author} {\bibfnamefont {S.~A.}\ \bibnamefont
			{Teukolsky}}, \bibinfo {author} {\bibfnamefont {W.~T.}\ \bibnamefont
			{Vetterling}},\ and\ \bibinfo {author} {\bibfnamefont {B.~P.}\ \bibnamefont
			{Flannery}},\ }\href@noop {} {\emph {\bibinfo {title} {Numerical Recipes 3rd
				Edition: The Art of Scientific Computing}}},\ \bibinfo {edition} {3rd}\ ed.\
	(\bibinfo  {publisher} {Cambridge University Press},\ \bibinfo {address}
	{USA},\ \bibinfo {year} {2007})\BibitemShut {NoStop}%
	\bibitem [{\citenamefont {Lee}(2008)}]{Lee2008}%
	\BibitemOpen
	\bibfield  {author} {\bibinfo {author} {\bibfnamefont {M.~J.}\ \bibnamefont
			{Lee}},\ }\bibfield  {title} {\bibinfo {title} {Pseudo-random-number
			generators and the square site percolation threshold},\ }\href
	{https://doi.org/10.1103/PhysRevE.78.031131} {\bibfield  {journal} {\bibinfo
			{journal} {Phys. Rev. E}\ }\textbf {\bibinfo {volume} {78}},\ \bibinfo
		{pages} {031131} (\bibinfo {year} {2008})}\BibitemShut {NoStop}%
	\bibitem [{\citenamefont {Mertens}\ and\ \citenamefont
		{Moore}(2012)}]{Mertens2012}%
	\BibitemOpen
	\bibfield  {author} {\bibinfo {author} {\bibfnamefont {S.}~\bibnamefont
			{Mertens}}\ and\ \bibinfo {author} {\bibfnamefont {C.}~\bibnamefont
			{Moore}},\ }\bibfield  {title} {\bibinfo {title} {Continuum percolation
			thresholds in two dimensions},\ }\href
	{https://doi.org/10.1103/PhysRevE.86.061109} {\bibfield  {journal} {\bibinfo
			{journal} {Phys. Rev. E}\ }\textbf {\bibinfo {volume} {86}},\ \bibinfo
		{pages} {061109} (\bibinfo {year} {2012})}\BibitemShut {NoStop}%
	\bibitem [{\citenamefont {Hovi}\ and\ \citenamefont
		{Aharony}(1996)}]{Hovi1996}%
	\BibitemOpen
	\bibfield  {author} {\bibinfo {author} {\bibfnamefont {J.-P.}\ \bibnamefont
			{Hovi}}\ and\ \bibinfo {author} {\bibfnamefont {A.}~\bibnamefont {Aharony}},\
	}\bibfield  {title} {\bibinfo {title} {Scaling and universality in the
			spanning probability for percolation},\ }\href
	{https://doi.org/10.1103/PhysRevE.53.235} {\bibfield  {journal} {\bibinfo
			{journal} {Phys. Rev. E}\ }\textbf {\bibinfo {volume} {53}},\ \bibinfo
		{pages} {235} (\bibinfo {year} {1996})}\BibitemShut {NoStop}%
	\bibitem [{\citenamefont {Yonezawa}\ \emph {et~al.}(1989)\citenamefont
		{Yonezawa}, \citenamefont {Sakamoto},\ and\ \citenamefont
		{Hori}}]{Yonezawa1989I}%
	\BibitemOpen
	\bibfield  {author} {\bibinfo {author} {\bibfnamefont {F.}~\bibnamefont
			{Yonezawa}}, \bibinfo {author} {\bibfnamefont {S.}~\bibnamefont {Sakamoto}},\
		and\ \bibinfo {author} {\bibfnamefont {M.}~\bibnamefont {Hori}},\ }\bibfield
	{title} {\bibinfo {title} {Percolation in two-dimensional lattices. {I. A}
			technique for the estimation of thresholds},\ }\href
	{https://doi.org/10.1103/PhysRevB.40.636} {\bibfield  {journal} {\bibinfo
			{journal} {Phys. Rev. B}\ }\textbf {\bibinfo {volume} {40}},\ \bibinfo
		{pages} {636} (\bibinfo {year} {1989})}\BibitemShut {NoStop}%
	\bibitem [{\citenamefont {Haji-Akbari}\ and\ \citenamefont
		{Ziff}(2009)}]{Haji-Akbari2009}%
	\BibitemOpen
	\bibfield  {author} {\bibinfo {author} {\bibfnamefont {A.}~\bibnamefont
			{Haji-Akbari}}\ and\ \bibinfo {author} {\bibfnamefont {R.~M.}\ \bibnamefont
			{Ziff}},\ }\bibfield  {title} {\bibinfo {title} {Percolation in networks with
			voids and bottlenecks},\ }\href {https://doi.org/10.1103/PhysRevE.79.021118}
	{\bibfield  {journal} {\bibinfo  {journal} {Phys. Rev. E}\ }\textbf {\bibinfo
			{volume} {79}},\ \bibinfo {pages} {021118} (\bibinfo {year}
		{2009})}\BibitemShut {NoStop}%
	\bibitem [{\citenamefont {Haji-Akbari}\ \emph {et~al.}(2015)\citenamefont
		{Haji-Akbari}, \citenamefont {Haji-Akbari},\ and\ \citenamefont
		{Ziff}}]{Haji-Akbari2015}%
	\BibitemOpen
	\bibfield  {author} {\bibinfo {author} {\bibfnamefont {A.}~\bibnamefont
			{Haji-Akbari}}, \bibinfo {author} {\bibfnamefont {N.}~\bibnamefont
			{Haji-Akbari}},\ and\ \bibinfo {author} {\bibfnamefont {R.~M.}\ \bibnamefont
			{Ziff}},\ }\bibfield  {title} {\bibinfo {title} {Dimer covering and
			percolation frustration},\ }\href
	{https://doi.org/10.1103/PhysRevE.92.032134} {\bibfield  {journal} {\bibinfo
			{journal} {Phys. Rev. E}\ }\textbf {\bibinfo {volume} {92}},\ \bibinfo
		{pages} {032134} (\bibinfo {year} {2015})}\BibitemShut {NoStop}%
	\bibitem [{\citenamefont {Gori}\ and\ \citenamefont
		{Trombettoni}(2015)}]{Gori2015}%
	\BibitemOpen
	\bibfield  {author} {\bibinfo {author} {\bibfnamefont {G.}~\bibnamefont
			{Gori}}\ and\ \bibinfo {author} {\bibfnamefont {A.}~\bibnamefont
			{Trombettoni}},\ }\bibfield  {title} {\bibinfo {title} {Conformal invariance
			in three dimensional percolation},\ }\href
	{https://doi.org/10.1088/1742-5468/2015/07/p07014} {\bibfield  {journal}
		{\bibinfo  {journal} {Journal of Statistical Mechanics: Theory and
				Experiment}\ }\textbf {\bibinfo {volume} {2015}},\ \bibinfo {pages} {P07014}
		(\bibinfo {year} {2015})}\BibitemShut {NoStop}%
\end{thebibliography}

%

\end{document}